\documentclass[journal]{IEEEtran}

\usepackage{xcolor,soul,framed} 

\colorlet{shadecolor}{yellow}
\usepackage[pdftex]{graphicx}
\graphicspath{{../pdf/}{../jpeg/}}
\DeclareGraphicsExtensions{.pdf,.jpeg,.png}

\usepackage[cmex10]{amsmath}
\usepackage{array}
\usepackage{mdwmath}
\usepackage{mdwtab}
\usepackage{eqparbox}
\usepackage{url}
\usepackage{svg}
\usepackage{subcaption}
\usepackage{siunitx}
\usepackage{multirow}
\usepackage{multicol}
\usepackage[breaklinks]{hyperref}
\usepackage[switch]{lineno}
\begin{document}

\bstctlcite{IEEEexample:BSTcontrol}
    \title{Picosecond Synchronization of Photon Pairs through a Fiber Link between Fermilab and Argonne National Laboratories}
     

\author{Keshav Kapoor, Si Xie, Joaquin Chung, Raju Valivarthi, Cristi\'{a}n Pe\~{n}a, Lautaro Narv\'{a}ez, Neil Sinclair, Jason P. Allmaras, Andrew D. Beyer, Samantha I. Davis, Gabriel Fabre, George Iskander, Gregory S. Kanter, Rajkumar Kettimuthu, Boris Korzh, Prem Kumar, Nikolai Lauk, Andrew Mueller, Matthew Shaw, Panagiotis Spentzouris, Maria Spiropulu, Jordan M. Thomas, and Emma E. Wollman
\thanks{K. Kapoor, C. Pe\~{n}a and P. Spentzouris are with the Fermi National Accelerator Laboratory, Batavia, IL 60510, USA}%
\thanks{S. Xie is with the Fermi National Accelerator Laboratory, Batavia, IL 60510, USA, the Division of Physics, Mathematics and Astronomy and Alliance for Quantum Technologies (AQT), California Institute of Technology, Pasadena, CA 91125, USA}%
\thanks{J. Chung, R. Kettimuthu are with the Argonne National Laboratory, Lemont, IL 60439, USA}%
\thanks{G. Iskander is with the Argonne National Laboratory, Lemont, IL 60439, USA and the University of Chicago, Physics Department, Chicago, IL 60637, USA}
\thanks{R. Valivarthi, L. Narv\`{a}ez, S.I. Davis, N. Lauk, G. Fabre, and M. Spiropulu are with the Division of Physics, Mathematics and Astronomy and Alliance for Quantum Technologies (AQT), California Institute of Technology, Pasadena, CA 91125, USA}%
\thanks{N. Sinclair is with the John A. Paulson School of Engineering and Applied Sciences, Harvard University, Cambridge, MA 02138, USA, the Division of Physics, Mathematics and Astronomy and Alliance for Quantum Technologies (AQT), California Institute of Technology, Pasadena, CA 91125, USA}%
\thanks{J.P. Allmaras, A.D. Beyer, B. Korzh, M. Shaw and E.E. Wollman are with the Jet Propulsion Laboratory, California Institute of Technology, Pasadena, CA 91109, USA}%
\thanks{G.S. Kanter is with the Center for Photonic Communication and Computing, Department of Electrical and Computer Engineering, McCormick School of Engineering and Applied Science, Northwestern University, 2145 Sheridan Road, Evanston, IL 60208, USA and NuCrypt LLC, 1460 Renaissance Dr \#205, Park Ridge, IL 60068, USA}%
\thanks{P. Kumar is with the Center for Photonic Communication and Computing, Department of Electrical and Computer Engineering, McCormick School of Engineering and Applied Science, Northwestern University, 2145 Sheridan Road, Evanston, IL 60208, USA and Department of Physics and Astronomy, Northwestern University, 2145 Sheridan Road, Evanston, IL 60208, USA}%
\thanks{A. Mueller is with the Division of Engineering and Applied Science, the Alliance for Quantum Technologies and the Jet Propulsion Laboratory,
California Institute of Technology, Pasadena, CA 91125, USA}%
\thanks{J.M. Thomas is with the Center for Photonic Communication and Computing, Department of Electrical and Computer Engineering, McCormick School of Engineering and Applied Science, Northwestern University, 2145 Sheridan Road, Evanston, IL 60208, USA}%
}



\maketitle

\begin{abstract}
We demonstrate a three-node quantum network for C-band photon pairs using 2 pairs of 59 km of deployed fiber between Fermi and Argonne National Laboratories.
The C-band pairs are directed to nodes using a standard telecommunication switch and synchronized to picosecond-scale timing resolution using a coexisting O- or L-band optical clock distribution system.
We measure a reduction of coincidence-to-accidental ratio (CAR) of the C-band pairs from 51 $\pm$ 2 to 5.3 $\pm$ 0.4 due to Raman scattering of the O-band clock pulses.
Despite this reduction, the CAR is nevertheless suitable for quantum networks.
\end{abstract}

\begin{IEEEkeywords}
quantum network, quantum communication, quantum-classical coexisting, clock distribution, fiber optics, photon pair, C-band, L-band, O-band, Raman noise
\end{IEEEkeywords}

%
\IEEEpeerreviewmaketitle

\section{Introduction}
Research quantum networks based on photon generation, interference, and detection, can be deployed at metropolitan distances.
This is due to their simplicity and reconfigurability, for example to multi- and dynamic-node topologies, and also  their ability to scale using off-the-shelf devices.
In addition, photonic quantum states can be generated and detected at high clock rates. In the future when quantum repeater devices such as quantum memories, typically narrowband~\cite{valivarthi2020teleportation},~\cite{valivarthi2016quantum},~\cite{alshowkan_2022},~\cite{chung_2021} evolve, they can be integrated in such high rate networks.
Since deployed networks unavoidably use channels that are perturbed by the environment, such as fiber optics which undergo thermal expansion, high rate networks require tracking of photons for high fidelity operation.
This is often achieved by distributing optical clock pulses along with the photons to establish a common time reference among the nodes.
These clock pulses account for any variations in travel time of the photons carrying quantum information.
Such approach has been used for several networking demonstrations \cite{tanaka08}, \cite{lessing_2017}, \cite{chen_2017}, \cite{wu_2019}. As the clock signals are much stronger than the quantum signals, considerations must be taken to ensure that the clock distribution system does not introduce noise while maintaining the timing resolution performance required for network operations. For instance, clock pulses are often at lower energies than the photons to minimally introduce noise by off-resonant scattering effects in the fiber, such as a Stokes field \cite{thomas_2021}. Despite this, we have previously demonstrated a picosecond-resolution clock distribution system for C-band photon pairs in a laboratory setting using an O-band clock, finding tolerable noise effects introduced by the clock distribution system \cite{Valivarthi:2022vni}. 
In this work, we present a demonstration of picosecond-scale synchronization of C-band photon pairs over deployed fiber optic cables connecting three remote nodes by distributing the clock pulses in either the telecommunication standard O- or L-bands.
Two of the nodes are located in the Fermi National Accelerator Laboratory (FNAL) FQNET/IEQNET site. These include the central node at the Feynman Computing Center (FNAL-FCC) and one of the end nodes at the D0 Assembly Building (FNAL-DAB), and are stationed 2 km apart.

The third node is located at the Argonne National Laboratory (ANL) IEQNET site.
This third node is referred to as ANL, and is 57 km away from FNAL-FCC. 
Each of these nodes are connected with pairs of fibers.
The central node at FNAL-FCC contains a photon pair source as well as a commercial all-optical telecom switch from Polatis and the end nodes at FNAL-DAB and ANL contain the superconducting nanowire single photon detectors (SNSPDs).
The quantum-correlated photon pairs are split, and each photon in the pair is sent into one of these 2~km-2~km and 57~km-57~km fiber pairs linking the nodes, corresponding to total transmission distances of the quantum states of 4~km and 114~km, with 2~km and 57~km of shared quantum-classical coexistence with a clock signal.  

Specifically, we study Raman scattering along each path, finding Raman coefficients consistent with previous work~\cite{frohlich_2015}.
We also use the clock distribution system of ref.~\cite{Valivarthi:2022vni} to demonstrate picosecond level synchronization of photon pairs between the nodes. 
Finally, we measure the noise introduced to the quantum information due to quantum-classical coexistence signals distributed between nodes in our network and find that in the FNAL-FCC to ANL link, the O- and L-band clocks reduce the coincidence-to-accidental ratios ($\mathrm{CAR}$) from $51 \pm 2$ to $5.3 \pm 0.4$ and $2.6 \pm 0.3$, respectively.
These measurements demonstrate that our research prototype network is suitable for point-to-point schemes and for two-photon interference-based teleportation protocols, representing a notable milestone towards establishing a national research quantum internet between the U.S. Department of Energy laboratories as envisioned in DOE's Quantum Internet Blueprint \cite{QIB:DOE2020-21}.

\section{Setup}

\begin{figure*}[h!]
\centering
  \includegraphics[width=\textwidth]{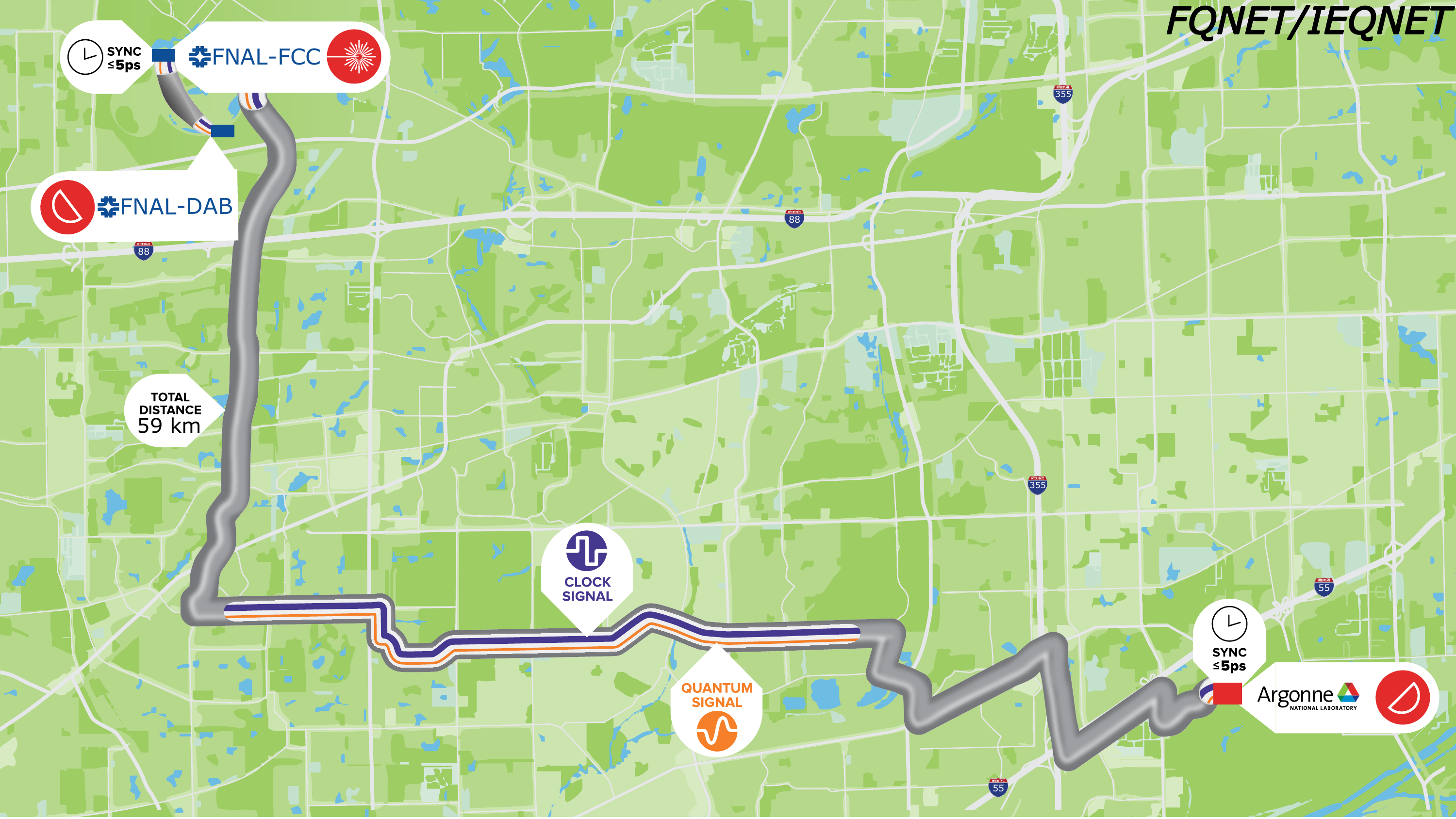}\\
  \caption{This image depicts the separation of the nodes in our network. FNAL-FCC and FNAL-DAB are connected with 2~km of dark fiber and FNAL-FCC and ANL are connected with 57~km of dark fiber. We keep our master clock at FNAL-FCC, and distribute the signal to FNAL-DAB and ANL, choosing the path via an optical switch located at FNAL-FCC. The FNAL nodes are depicted by the blue rectangles and the ANL node is depicted by the red rectangle.
  }
  \label{fig:map}
\end{figure*}

To characterize the noise introduced by the clocks in the network links for transmission of quantum information we make use of photon pair sources built from fiber-based, off-the-shelf, components. 
The optical switch located at FNAL-FCC is used to centrally route signals between the optical fiber connections between the three network nodes through a web-based control software interface \cite{polatis,ieqnet_design}. 
A map of the three-node quantum network is shown in Fig.~\ref{fig:map} along with the approximate paths of the optical fiber links between the nodes.

\subsection{Entangled Photon Pair Source}
\label{sec:PairSource}
Our photon pair source is situated at FNAL-FCC. Using a commercial arbitrary waveform generator (AWG), short radio-frequency (RF) pulses with widths of 80~ps and separated by 5~ns are generated, amplified, and used as the input to a fiber-coupled Mach-Zehnder Modulator (MZM) \cite{tektronix}, \cite{ixblue}.
Light at 1536~nm wavelength produced by a continuous wave fiber-coupled laser is directed into the MZM to produce pulsed light.
The pulsed light is directed into an erbium-doped fiber amplifier (EDFA) and then sent through a periodically poled lithium niobate (PPLN) waveguide to upconvert the 1536~nm light to 768~nm.
A band-pass filter is used to remove any residual 1536~nm light.
A second PPLN waveguide takes the 768~nm light as input to produce time-correlated photon pairs at the original wavelength of 1536~nm through Type-II spontaneous parametric down conversion process (SPDC).
A fiber-based polarizing beam splitter separates the photon pair into individual photons, one of which is directed to a dense wavelength division multiplexer (DWDM) and multiplexed with the clock signals described in the next subsection. 
The multiplexed quantum and classical channels are sent to the optical switch located at FNAL-FCC, which routes their path to FNAL-DAB or ANL.
The other photon of the pair is sent directly, using a dedicated fiber, to the optical switch and routed to the same node.

\begin{figure*}[h!]
  \centering
    \includegraphics[width=\textwidth]{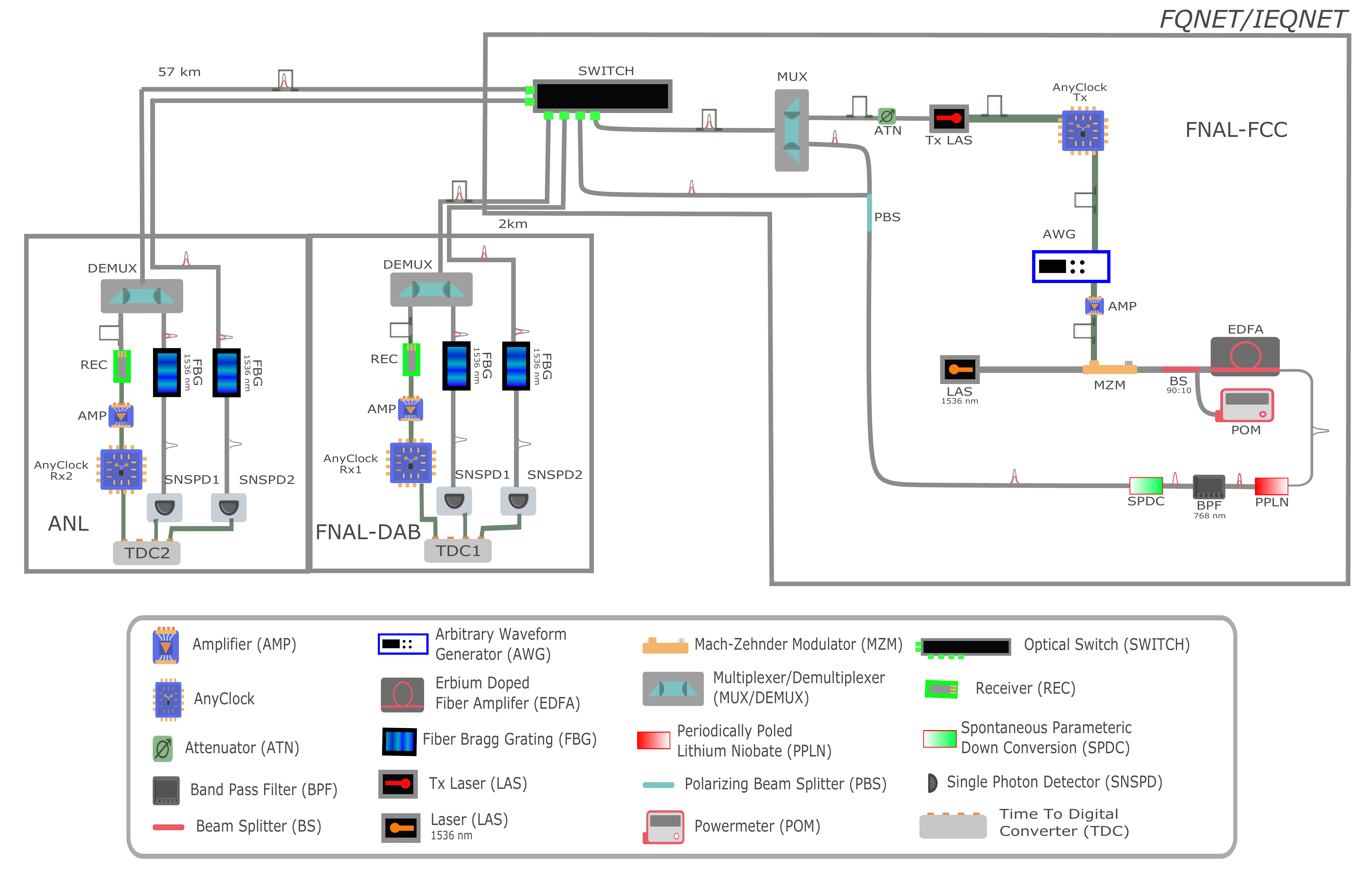}
  \caption{Schematic for the network we characterize. The square pulses represent the clock signal while the grey and red Gaussian-shaped pulses represent the quantum light and its second harmonic (768~nm), respectively. The photon pairs are produced at FNAL-FCC and routed either 2~km away to FNAL-DAB or 57~km away to ANL through software provided with the optical switch \cite{polatis}.}
  \label{fig:setup}
\end{figure*}

At the FNAL-DAB and ANL nodes, the combined quantum and clock channels are sent through a DWDM de-multiplexer (DEMUX) to separate them. 
The quantum channel is filtered by two or three additional DEMUX's and then a 4~GHz bandwidth fiber Bragg grating (FBG) filter to isolate the 1536~nm quantum frequency channel to within a wavelength of 0.03~nm. 
These filters reduce the effects of dispersion and improves the indistinguishability of the photons.
The FNAL-DAB and ANL nodes are equipped with two superconducting nanowire single photon detectors (SNSPD) to detect the incoming photons with timing jitter (resolution) of less than 50~ps.
The SNSPD signals are digitized by commercial time-taggers with timing jitter below 10~ps.
A schematic of our network is shown in Fig~\ref{fig:setup}.

\subsection{Clock Distribution and Synchronization}


To synchronize the clocks at the FNAL-DAB and ANL nodes with the clock of the central FNAL-FCC node, we utilize the system described in Ref.~\cite{Valivarthi:2022vni}. 
The transmitter (Tx) module consists of an O-~(1310~nm) or L-band (1610~nm) laser diode that is bias switched via 2.5~ns-duration pulses generated by a 200~MHz voltage oscillator.
This Tx is used to distribute clock signals from the central FNAL-FCC node to the FNAL-DAB and ANL nodes on the same fiber that is used to transmit the single photons.
At the FNAL-DAB and ANL nodes, the receiver (Rx) consists of a 200~MHz-bandwidth photodetector that generates electrical pulses amplified by a custom-designed 15~dB amplifier~\cite{Valivarthi:2022vni}.
These pulses adjust the phase of 200~MHz voltage oscillator clocks at the FNAL-DAB and ANL nodes which are used as time references.
The time taggers record the time difference between the electrical signal pulses generated by the SNSPDs and the reference clock signal.
The time differences are logged using a scalable data acquisition and monitoring system, enabling uninterrupted quantum network operations.

\section{Results}

We characterize the three-node quantum network by measuring the Raman scattering coefficient, timing jitter of the clock distribution system, and coincidence-to-accidental ratio (CAR) of photon pairs distributed from FNAL-FCC to FNAL-DAB and FNAL-FCC to ANL.

\subsection{Raman Scattering Coefficient}

We measure the Raman scattering coefficients between FNAL-FCC and ANL, and between FNAL-FCC and FNAL-DAB.
We obtain the Raman scattering coefficients by measuring the transmitted power of the O- or L-band clock pulses that are scattered into the C-band quantum channel at the end of each fiber.
The transmitted power $P_r$ of the Raman scattered photons is
\begin{equation}
    \eta_s P_r = P_0\beta \Delta \lambda \frac{e^{-\alpha_C L} - e^{-\alpha_{O,L} L}}{\alpha_{O,L} - \alpha_{C}},
    \label{eqn:beta}
\end{equation}
where $P_0$ is the average launch power of the O- or L-band optical clock pulses, $\Delta\lambda$ is the bandwidth of the narrowband filter, $L$ is the distance traveled along the optical fiber, $\alpha_C$ and $\alpha_{O,L}$ are the loss coefficients for the C-, O-, and L-bands, respectively, $\eta_s$ accounts for insertion losses in the DEMUXs and FBGs, as well as the efficiency of the SNSPD, and $\beta$ is the Raman scattering coefficient \cite{frohlich_2015}.
We send a launch power into the clock channel from the central FNAL-FCC node to the FNAL-DAB and ANL endnodes.
The filtered light is sent into the SNSPD to measure the photon rate and thus determine $P_r$.
The Raman scattering coefficient $\beta$ is calculated from equation~\ref{eqn:beta}.

\begin{figure}
  \begin{center}
  \includegraphics[width=3.5in]{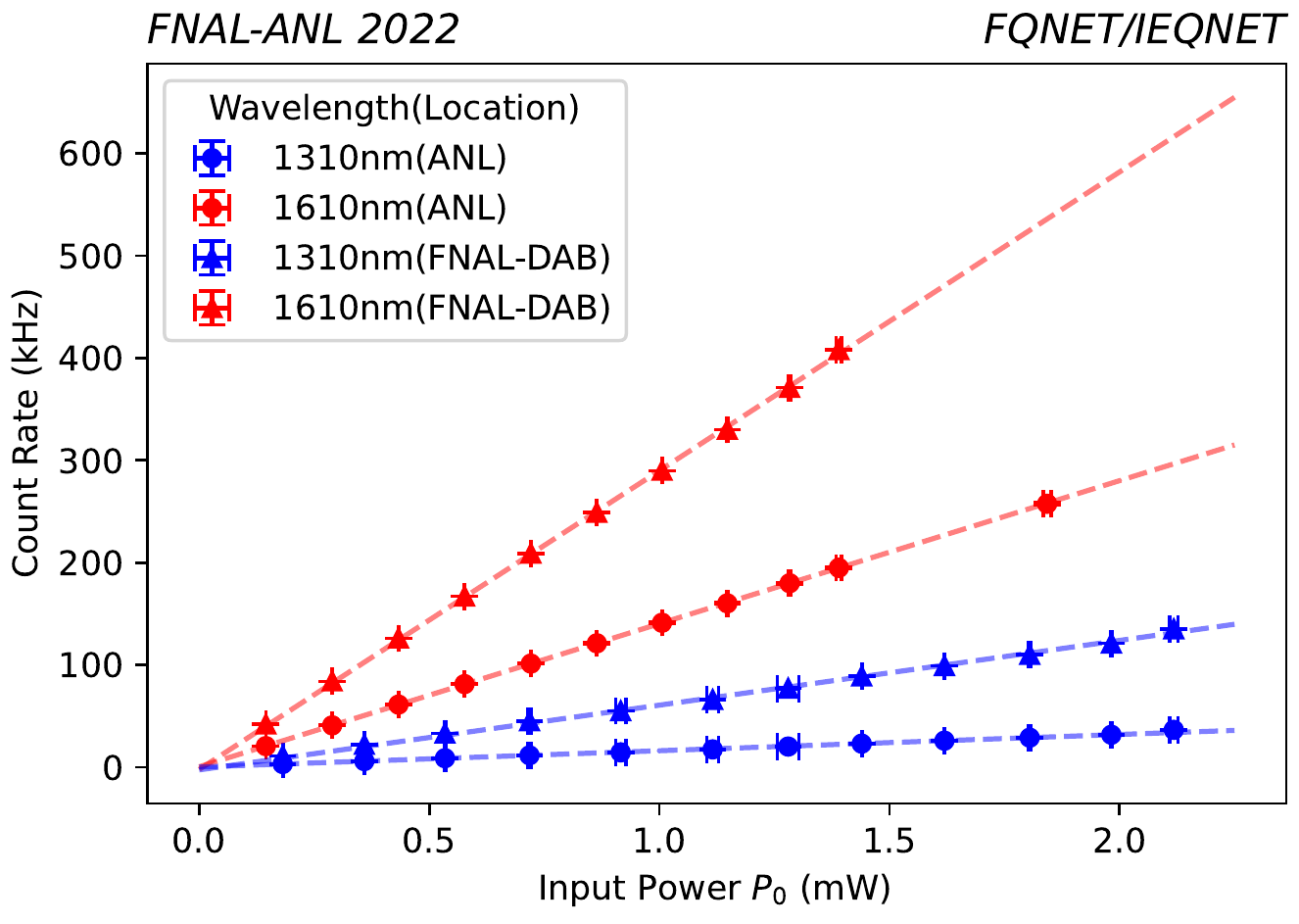}\\
  \caption{The measured count rate of Raman scattered light into the quantum channel C-band window are shown for the L- (blue) and O-band (red) clock signals.
  The measurements made at the FNAL-DAB and ANL end nodes from signals sent from the FNAL-FCC central node are shown in the triangular and circular data points, respectively. }
  \label{fig:rates}
  \end{center}
\end{figure}

The Raman scattered power for varied inputs into both fiber connections is shown in Fig. ~\ref{fig:rates}.
The measurements for the O-band and L-band clocks are shown in blue and red, respectively. 
Linear fits yield slopes of $P_r/P_0$, from which we derive the Raman scattering coefficients, $\beta$, for the fiber links from the central FNAL-FCC node to the FNAL-DAB and ANL end nodes.
The fiber length and loss coefficients between the nodes are obtained from independent optical time-domain reflectometry (OTDR) measurements, these values can be found in Table~\ref{table:beta}.
The insertion loss in the DEMUX and FBG filters are measured and used to determine $\eta_s$, which is $2.88$ for the O-band and $2.12$ for the L-band.
The resulting $\beta$ are shown in Table~\ref{table:beta}, for both end nodes and both types of clocks.

\begin{table}[]
\begin{center}
\begin{tabular}{| c | c | c |}
        \hline \multirow{2}{*}{Wavelength (nm)}
        & \multicolumn{2}{c|}{$\beta \left(nm^{-1} km^{-1}\right) \times 10^{-10}$}          \\
        \cline{2-3}
        & FNAL-DAB     & ANL       \\
        \hline
1610 & $33 \pm 3.0$    & $20.8\pm0.3$  \\
\hline
1310 & $10.5 \pm 0.3$ & $4.6 \pm 0.1$ \\
\hline 
& \multicolumn{2}{c|}{$\alpha \left(km^{-1}\right)$} \\
\hline
1610 & 0.5 & 0.084  \\
\hline
1310 & 0.55 & 0.099  \\
\hline
1536 & 0.44 & 0.076 \\
\hline
\end{tabular}
\caption{Raman scattering coefficients ($\beta$) and the power attenuation coefficients ($\alpha$) for different wavelengths for the fibers connecting FNAL-FCC to FNAL-DAB and FNAL-FCC to ANL.
\label{table:beta}
}
\end{center}
\end{table}

Based on the measured $\beta$ we predict the Raman scattered power for the L- and O-band clocks for varied fiber length based on the minimum launch power needed to synchronize our clock system at the given fiber length, shown in Figure~\ref{fig:gui}.
These results can be used to model the behavior for future network links and guide network design and optimization.

\begin{figure*}
  \begin{center}
  \begin{subfigure}[b]{0.5\textwidth}
    \begin{center}
        \includegraphics[width=\textwidth]{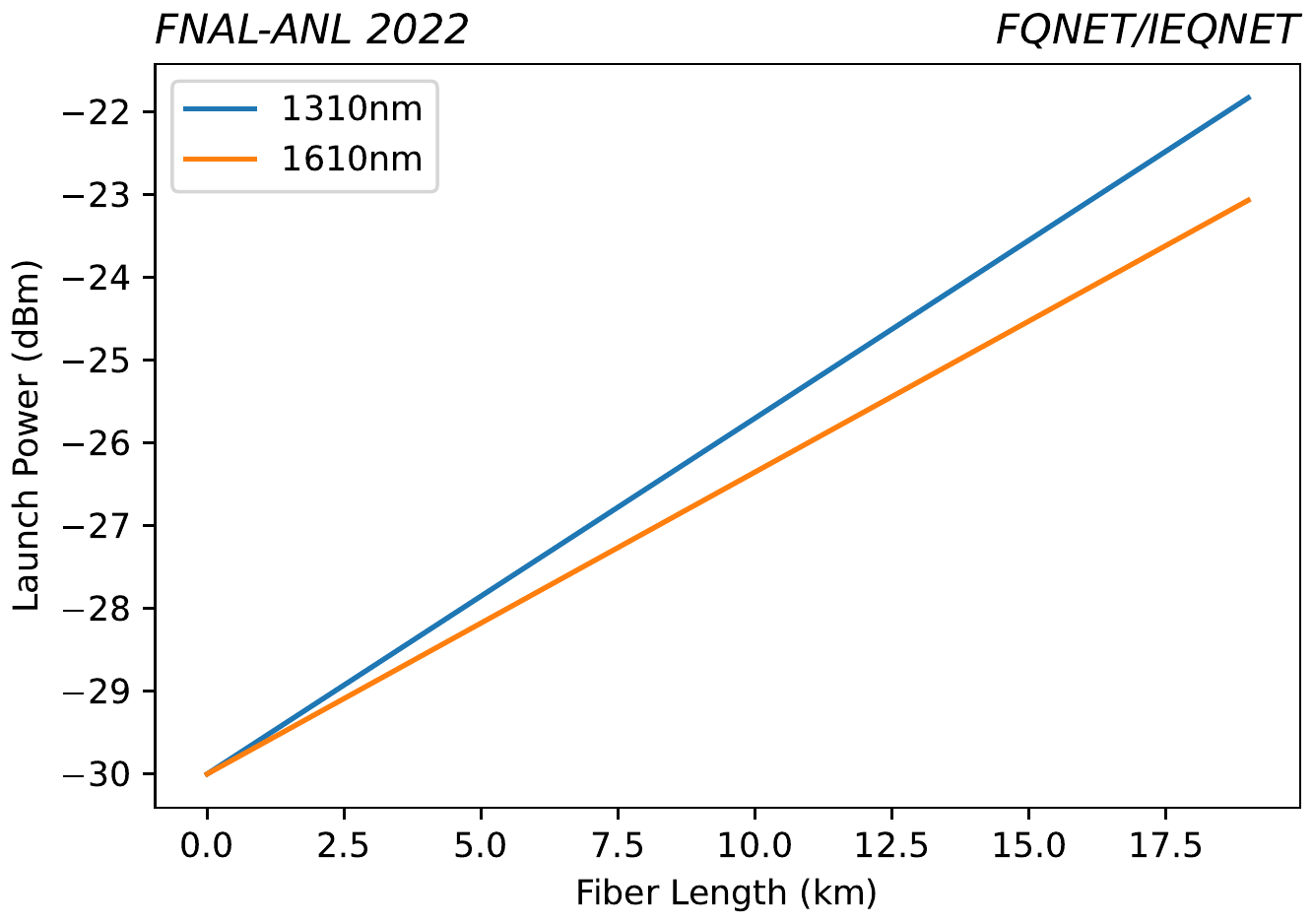}
    \end{center}
  \end{subfigure}
  \hspace{-0.3cm}
  \begin{subfigure}[b]{0.5\textwidth}
    \begin{center}
        \includegraphics[width=\textwidth]{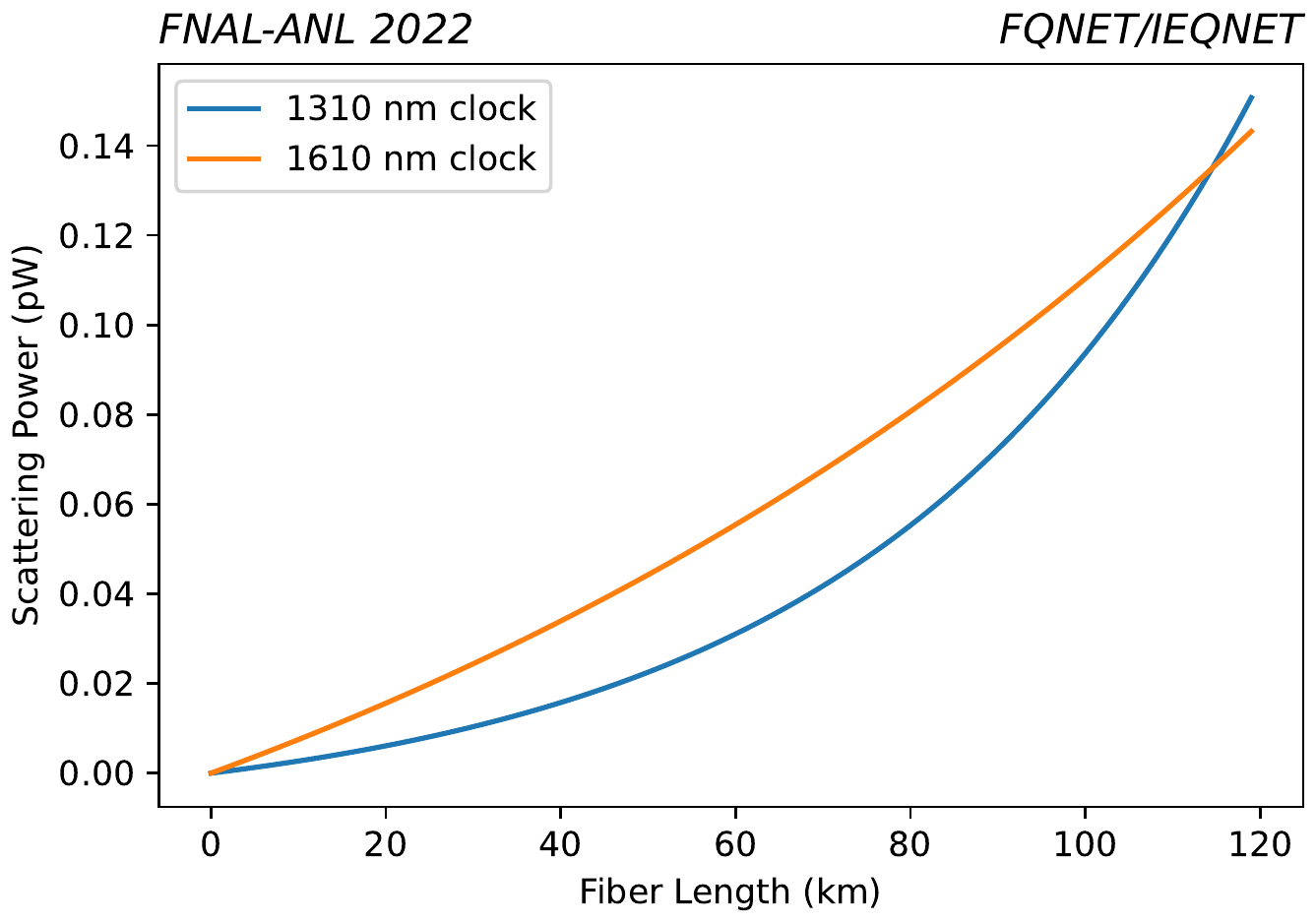}
    \end{center}
  \end{subfigure}
  \caption{(Left) The minimum launch powers, $P_0$, we need to synchronize the clock as a function of fiber length for the L- and O-bands, the solid orange and blue lines respectively. (Right) The power of the light Raman Scattered into the quantum channel from the L- and O-bands, the solid orange and blue lines respectively, as a function of fiber length. Both left and right plots use the values for $\beta$ and $\alpha$ corresponding to the FNAL-FCC to ANL link as shown in TABLE~\ref{table:beta}.}
  \label{fig:beta_lengths}
  \end{center}
\end{figure*}

\subsection{Clock Synchronization}

\begin{figure}
  \begin{center}
  \includegraphics[width=0.5\textwidth]{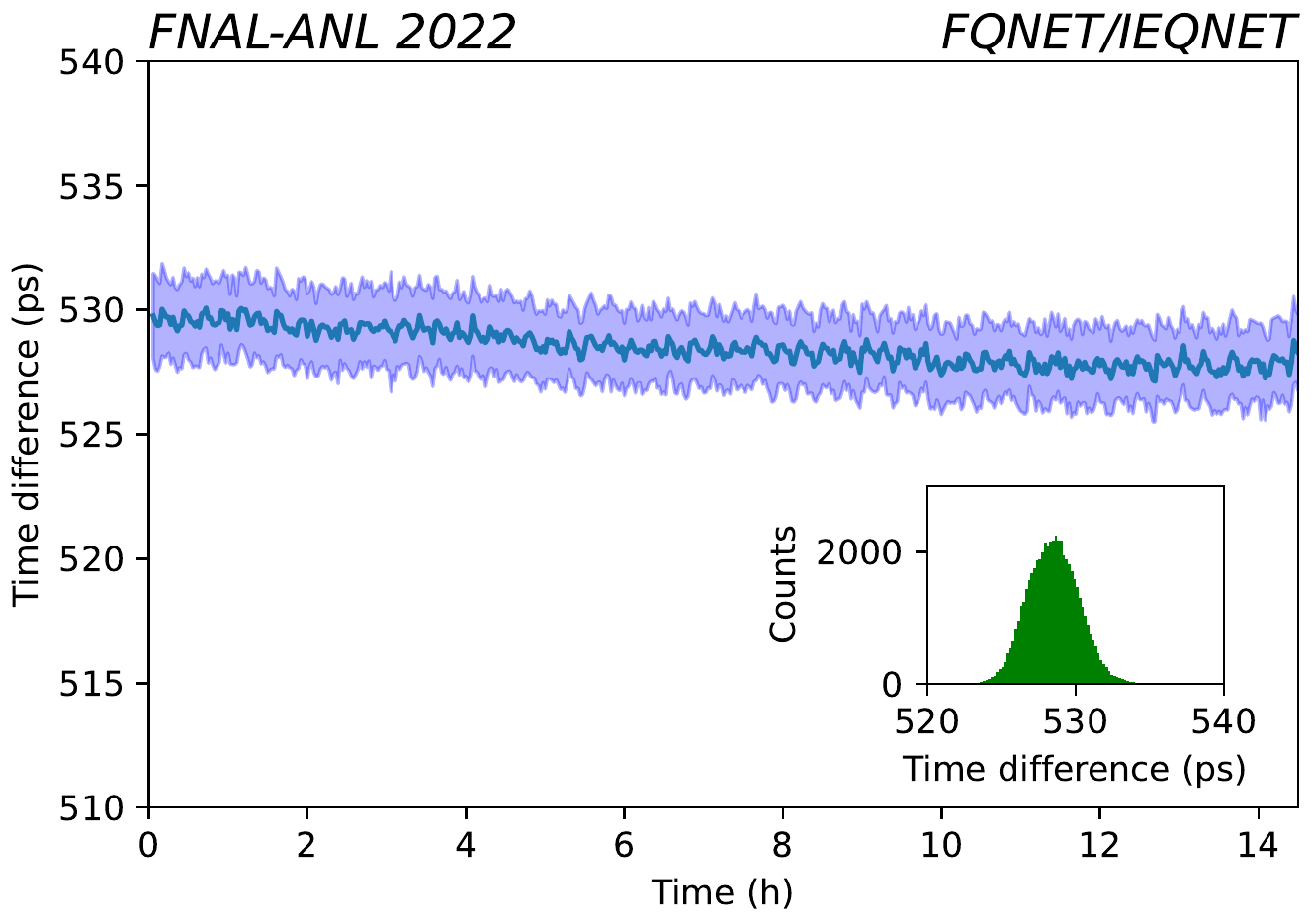}\\
  \caption{Variation of the time difference between the arrival of the reference and derived clock at ANL over 14 hours. The blue line is the average of the time difference every 100 seconds, showing the drift in the time difference of the two clocks. The blue shaded region is the RMS of the time difference during each of those 100 second intervals. Inset: histogram of the time difference indicates a timing jitter of 2.2 ps.}
  \label{fig:jitter}
  \end{center}
\end{figure}

We evaluate the drift and timing jitter of the clock synchronization system by monitoring the time difference between the synchronized local clock and a reference signal pulse produced by the same 1536~nm laser that is used for the entangled pair source. 
The reference signal pulse is modulated by the MZM that is driven by an AWG synchronized with the master clock at the FNAL-FCC central node, and subsequently amplified by the EDFA.
This reference signal is transmitted along a fiber parallel to the fiber used to distribute the master clock to the FNAL-DAB and ANL end nodes.
At the end nodes, the reference signal is detected by a fast 20~GHz photodetector and digitized by the time tagger.
The time difference between the synchronized local clock and the reference signal is collected over each second then plotted in a histogram and long term time drift is monitored through the mean difference, this drift as well as the representative time difference over a period of more than 14~hours can be seen in Figure~\ref{fig:jitter}.

We observe a drift in the mean of less than 5~ps and a total jitter of 2.2~ps.
This is much less than the 250~ps duration of our photons, which renders our system applicable for quantum networks including those that rely on two-photon interference \cite{valivarthi2020teleportation}.

\subsection{Noise Characterization}

We measure the noise introduced by the clock pulses by measuring the reduction of $\mathrm{CAR}$ of our time-correlated photon pairs when they are measured locally at the central FNAL-FCC node compared to when they are distributed over fiber with the clock pulses, where $\mathrm{CAR}=\mathrm{C}/\mathrm{A}$, where $\mathrm{C}$ is the coincidence detection rate of photons from the same pair and $\mathrm{A}$ is the coincidence detection rate of photons that are not from the same pair. 
Raman scattered photons will increase the detection of accidental occurrences thus further reducing the $\mathrm{CAR}$.
The CAR is a commonly used metric for evaluating non-classicality of two-photon fields \cite{Valivarthi:2022vni}.

From the central FNAL-FCC node, we send to the FNAL-DAB and ANL end nodes the O- and L-band clocks multiplexed along the same fiber that is used to send one of the photons from the correlated pair carrying the quantum signal as described in Section~\ref{sec:PairSource} above.
On a separate fiber parallel to the fiber carrying the multiplexed channel, we send the second photon of the correlated pair.
We send clock signals at 0.3~mW power for both the O- and L-bands to the FNAL-DAB end node, while for the ANL end node we send the O-band clock signal at 1.8~mW power and the L-band at 0.3~mW power.

In the absence of Raman scattered photons and dark counts, the $\mathrm{CAR}$ is equivalent to the cross-correlation $g^{(2)}(0)$ function, which quantifies the ratio of the detection of correlated photon pairs to non-correlated photon pairs resulting from more than one photon pair being produced at the pair source~\cite{loudon2000}.
We count all photon pair detection events within a 450~ps interval around the main coincidence peak to obtain $\mathrm{C}$, and take the average of the photon pair detection events within a 450~ps interval around each of the accidental peaks to determine $\mathrm{A}$. 
The measurement is made over a period of 5 minutes for the FNAL-DAB end node.
Due to higher losses from the longer distance resulting in lower coincidence rates, it takes 12 hours to make the analogous measurement for the ANL end node. 

\begin{figure}
  \begin{center}
  \includegraphics[width=0.5\textwidth]{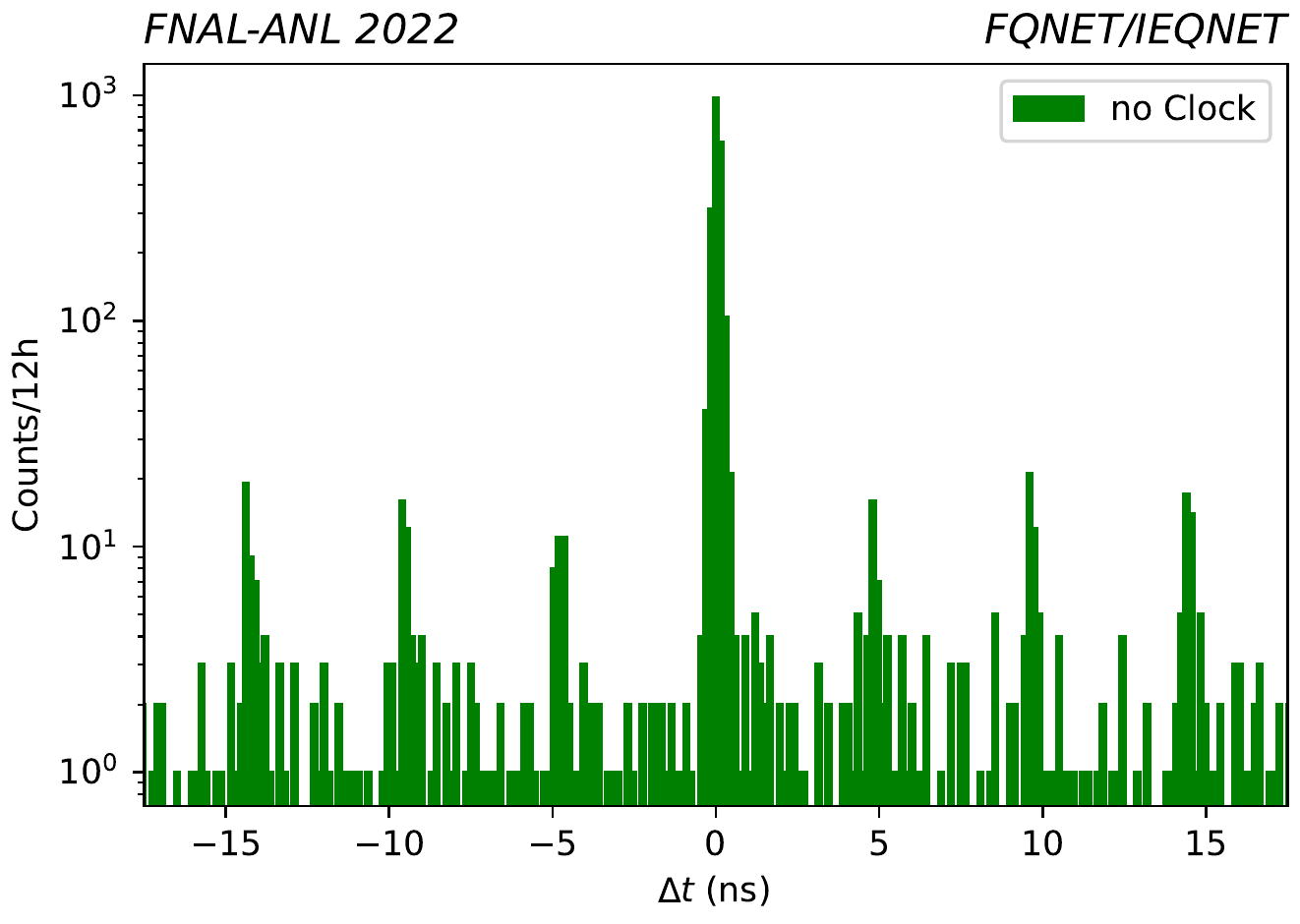}\\
  \caption{Coincidence histogram for the photon pairs sent to ANL from FNAL-FCC with the clock distribution disabled.}
  \label{fig:CAR}
  \end{center}
\end{figure}

In Figure~\ref{fig:CAR}, we show the time difference distribution for the two detected photons in a background-free scenario with the photon pairs sent from the central FNAL-FCC node to the ANL end node and no synchronization clock signal being sent along the same fiber.
The main coincidence peak is clearly visible at the center ($\Delta t = 0$), and the smaller peaks separated by 5~ns are the ``accidental" events corresponding to a detection of one photon from one pair and another photon from preceding or subsequent pairs. 
We measure a $\mathrm{CAR}$ of 51 $\pm$ 2 in the background-free scenario.

\begin{figure}
  \begin{center}
  \begin{subfigure}[b]{0.24\textwidth}
    \begin{center}
        \includegraphics[width=\textwidth]{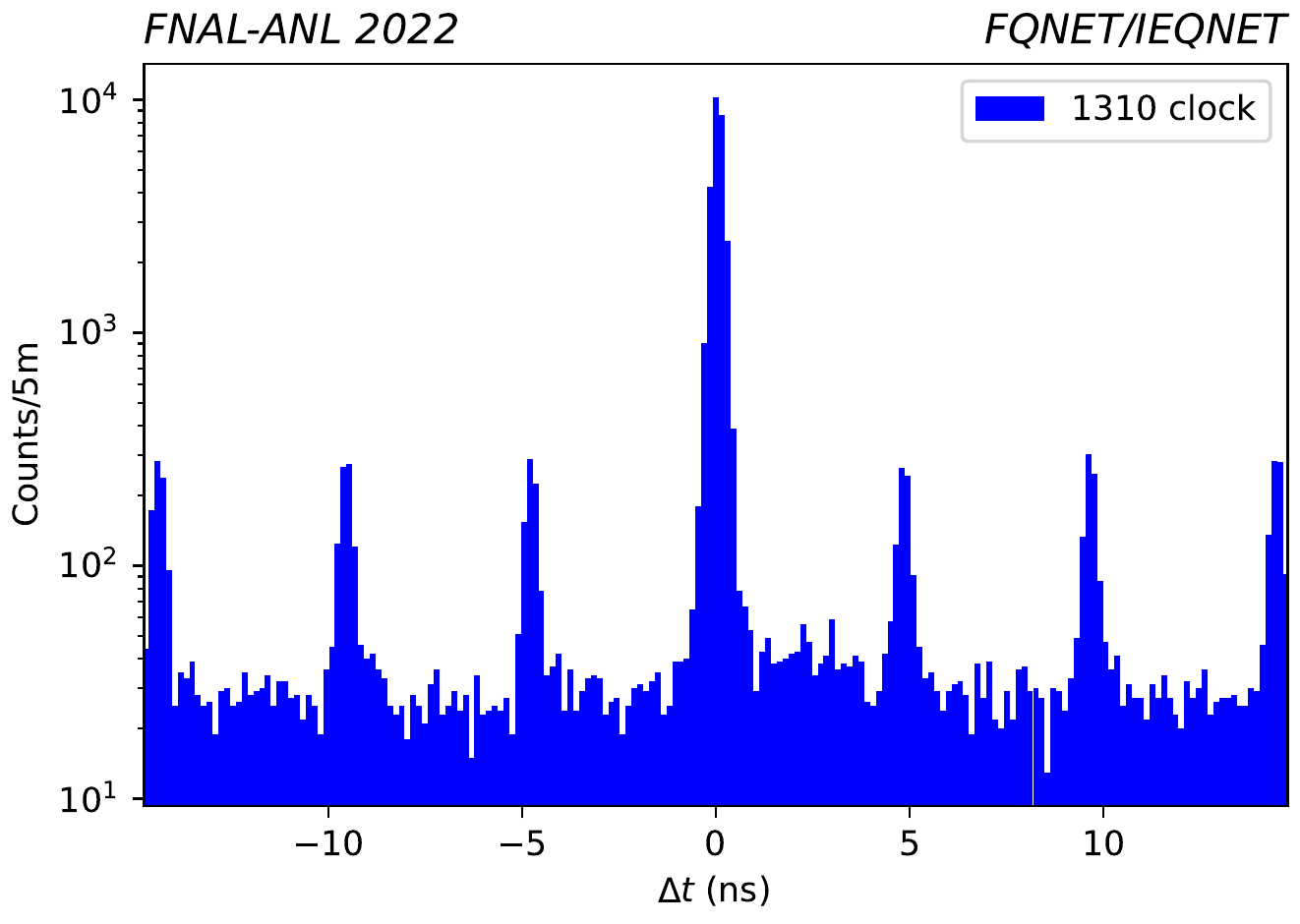}
    \end{center}
  \end{subfigure}
  \hspace{-0.3cm}
  \begin{subfigure}[b]{0.24\textwidth}
    \begin{center}
        \includegraphics[width=\textwidth]{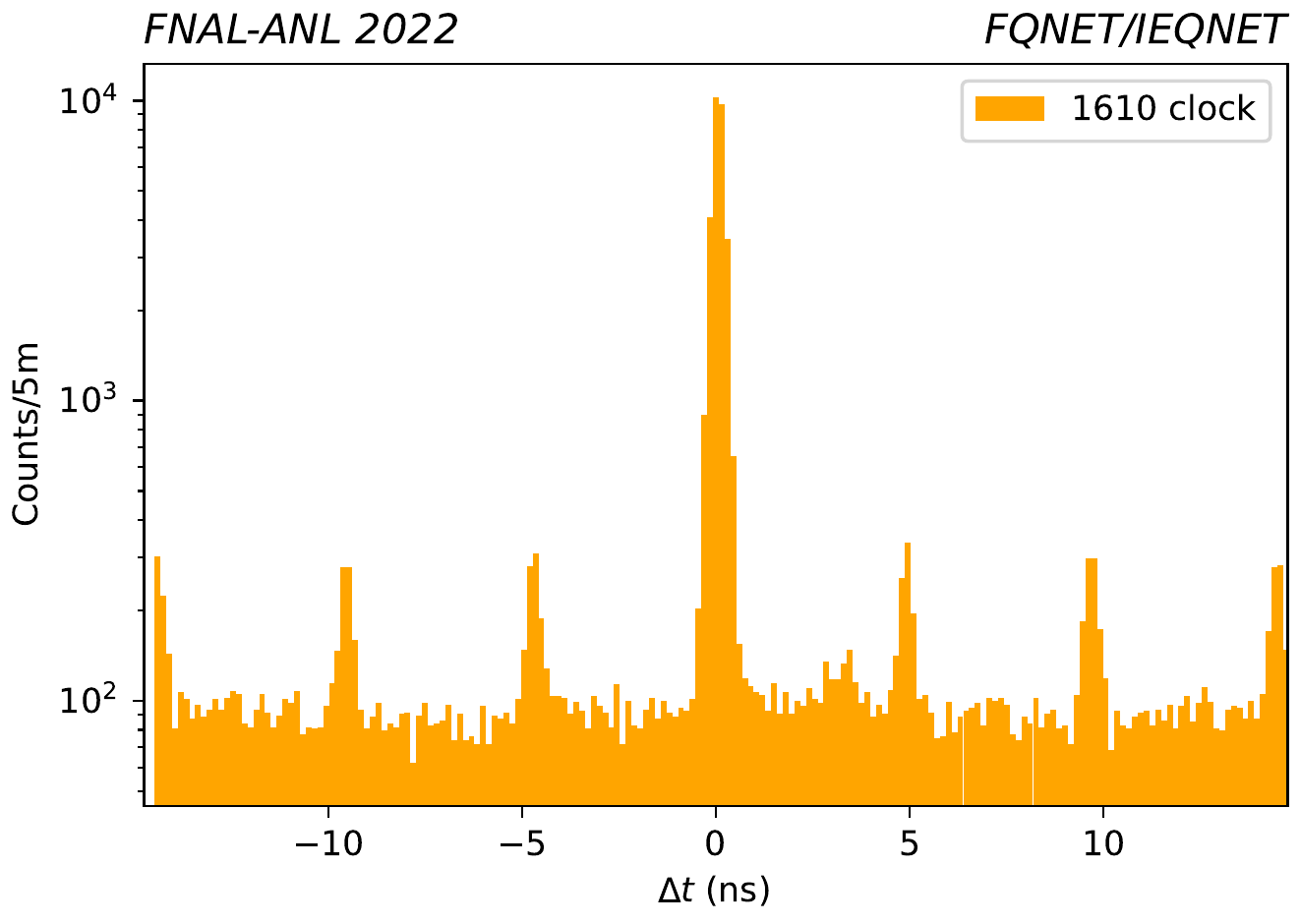}
    \end{center}
  \end{subfigure}
  \caption{(Left) coincidence histogram for the photon pairs sent to FNAL-DAB from FNAL-FCC with the 1310~nm clock distribution enabled. (Right) coincidence histogram for the photon pairs sent to FNAL-DAB from FNAL-FCC with the 1610~nm clock distribution enabled.
  \label{fig:CAR_D0}
  }
  \end{center}
\end{figure}

\begin{figure}
  \begin{center}
  \begin{subfigure}[b]{0.24\textwidth}
    \begin{center}
        \includegraphics[width=\textwidth]{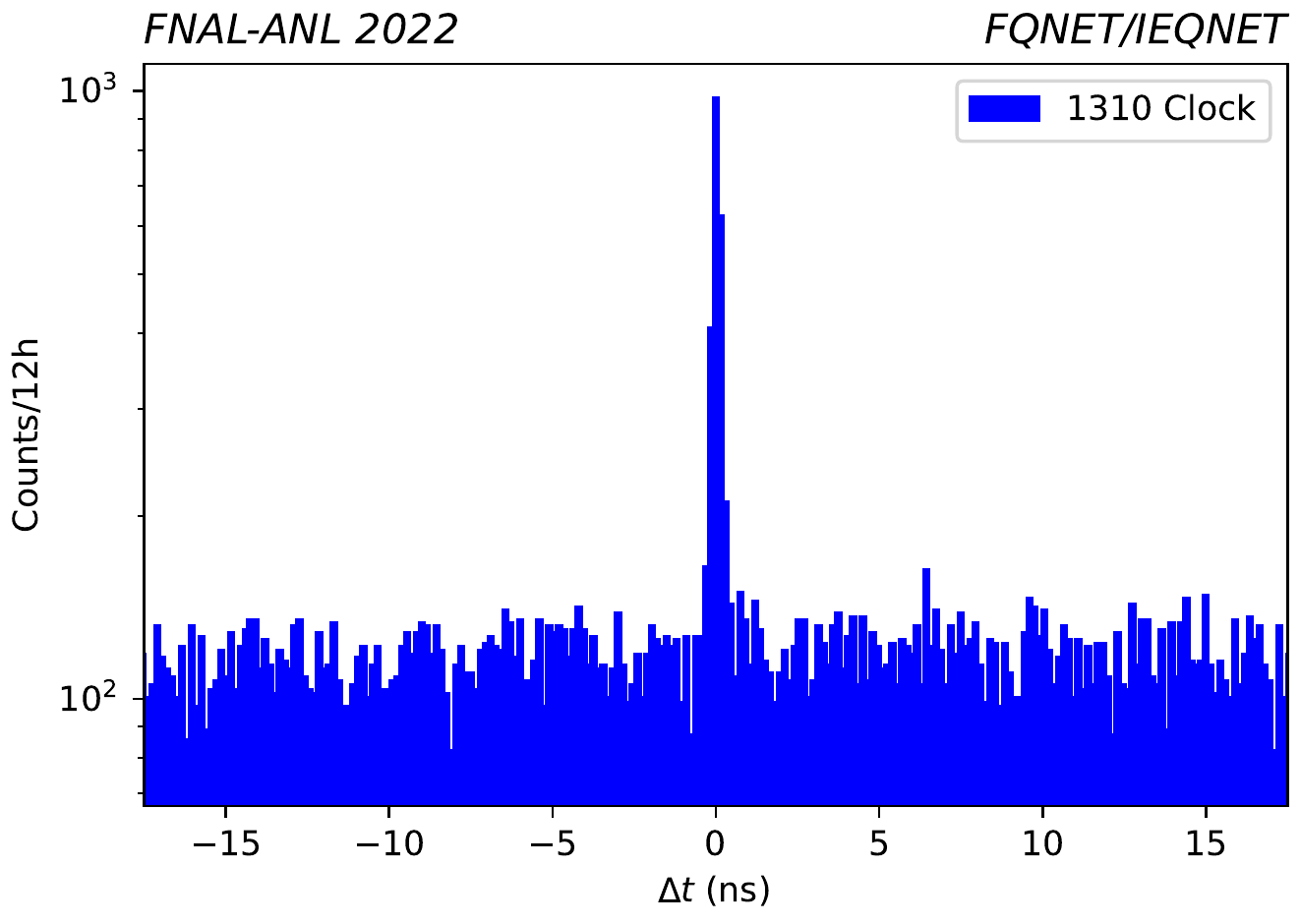}
    \end{center}
  \end{subfigure}
  \hspace{-0.3cm}
  \begin{subfigure}[b]{0.24\textwidth}
    \begin{center}
        \includegraphics[width=\textwidth]{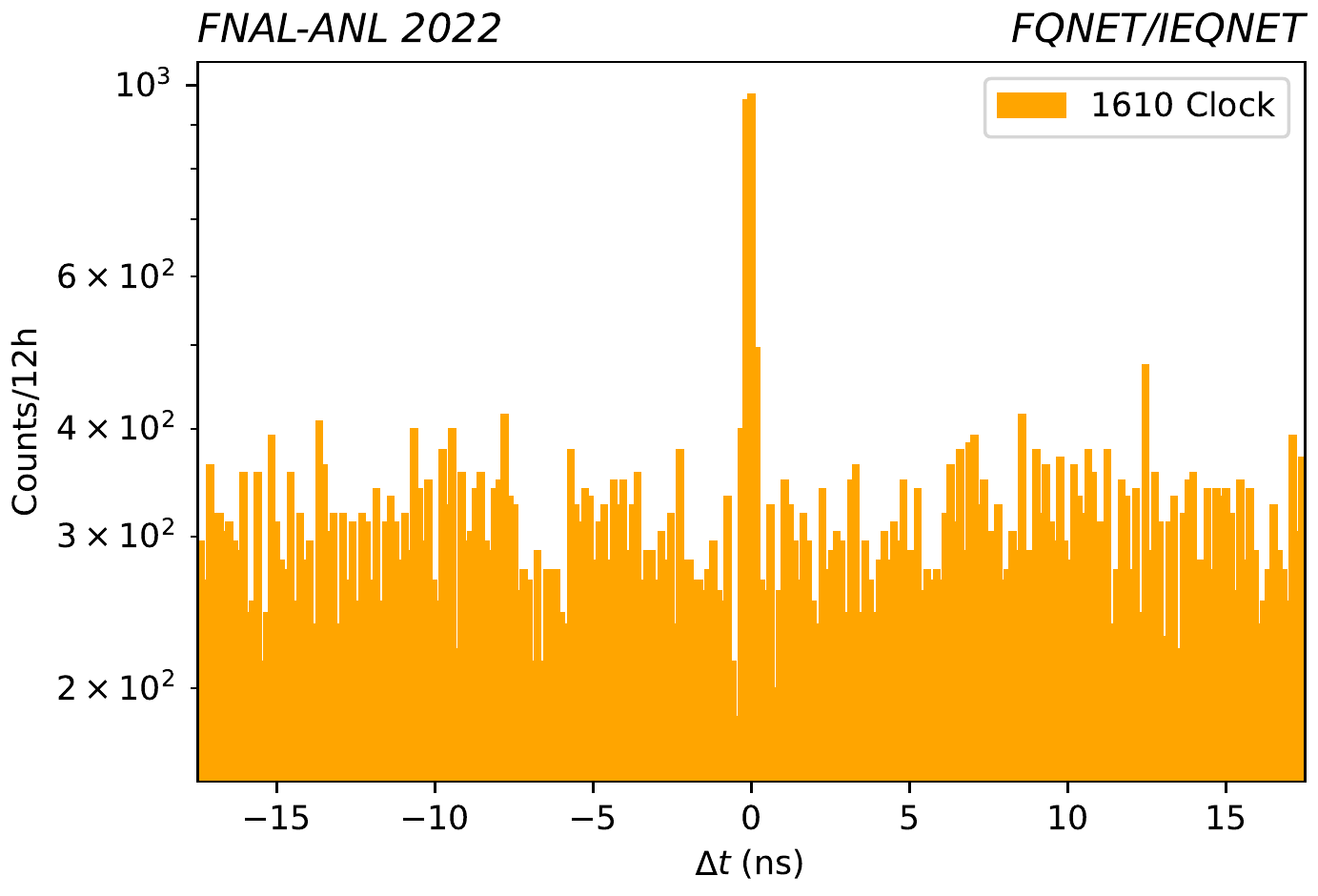}
    \end{center}
  \end{subfigure}
  \caption{(Left) coincidence histogram for the photon pairs sent to ANL from FNAL-FCC with the 1310~nm clock distribution enabled. (Right) coincidence histogram for the photon pairs sent to ANL from FNAL-FCC with the 1610~nm clock distribution enabled.
  \label{fig:CAR_ANL}
  }
  \end{center}
\end{figure}

Next, we sent the photon pairs to the FNAL-DAB and ANL end nodes with the O- and L-band clock synchronization enabled and multiplexed along the same fiber as one of the pairs. 
The time difference distributions of the two detected photons are shown in Figures~\ref{fig:CAR_D0}~and~\ref{fig:CAR_ANL} for the FNAL-DAB and ANL nodes, respectively. 
At the FNAL-DAB end node we measured a $\mathrm{CAR}$ of 35 $\pm$ 1 with the O-band clock, and 32 $\pm$ 1 with the L-band clock. 
At the ANL end node we measured a $\mathrm{CAR}$ of 5.3 $\pm$ 0.4 with the O-band clock, and 2.6 $\pm$ 0.3 with the L-band clock.
The width of the peaks in our measurements are limited by timing jitter in our detectors and readout electronics. 
These reductions in CAR are consistent with the model we explored in Figure~\ref{fig:beta_lengths}.
We see a reduction in CAR as we increase the length of the fiber, but we stay in the regime where the L-band clock introduces more noise than the O-band clock. 

\section{Discussion and outlook}
\label{sec:discussion}
We explored the practicality of the metropolitan-scale network for the purpose of transmitting quantum information.
We observe that a $\mathrm{CAR} > 2.0$, which is above the classical limit, is maintained in all the configurations we tested including using the 57~km installed fiber link between FNAL-FCC and ANL, which has increased losses due to fiber splicing and connectors along the real world fiber connection.
Comparing the idealized fiber attenuation loss coefficients in the C-band, our total losses correspond to a transmission of the quantum states over 105~km.
We showed that O-band clock signals distributed in coexistence with the quantum channel in the C-band is preferred over the L-band clock signals as it produces less Raman scattering into the C-band and results in less noise in the quantum channel.

We verified that reducing the power of the clock reduces the Raman-scattered background counts linearly.
This observation enables two future strategies to mitigate the effect of Raman-scattering to yield substantial improvement in the noise introduced to the quantum channel, quantified as $\mathrm{CAR}$. 
The first strategy is to reduce the duty cycle of the clock signal to 5--10\% (from the 50\% currently used) which will result in 5--10 times larger $\mathrm{CAR}$.
Further reduction of noise in our system can be achieved by using photodetectors with improved signal-to-noise ratio (SNR) so that the clock pulses can be recovered with the same precision at lower launch powers. An example of such devices are SNSPDs~\cite{korzh2020lowjitter} which we and others have already deployed in research quantum networking testbeds.

Additionally, we demonstrated a clock distribution system capable of synchronizing quantum nodes connected with up to 57~km of optical fiber with timing resolution better than 10~ps. The obtained timing resolution enables quantum sources capable of achieving a repetition rate of the order of 10~GHz, significantly improving the detection rates reported in this study. Furthermore, the clock distribution system can be easily scaled to tens of nodes with minimal modifications to the current design.
In addition, the CAR could be improved 
if the photon pair source was operated at a wavelength that is blue shifted from the clock, such as the O-band, while operating the clock in the C-band. Nonetheless, the C-band features the lowest loss in standard single-mode fiber used for long-haul networks (0.18~dB/km), while the O-band is a commonly used channel, and provides an operating wavelength that matches those of readily available off-the-shelf telecommunication components (e.g., lasers, modulators, MUX/DEMUXs, detectors, etc.).

Our three-node quantum network and accompanying synchronization system provides a benchmark on the role of noise in quantum networking and is a notable step towards practical, high-rate, classical-quantum co-existing networks.

\section*{Acknowledgment}
This work is partially funded by the Department of Energy Advanced Scientific Computing Research Transparent Optical Quantum Networks for Distributed Science program, IEQNET Grant~\cite{ASCR:IEQNET2019}. 
This work is also partially funded by the Department of Energy BES HEADS-QON Grant No. DE-SC0020376 on transduction relevant research for future quantum teleportation systems and communications.  R.V., N.L., L.N., N.S., G.F., M.S. and S.X. acknowledge partial and S.D. full support from the Alliance for Quantum Technologies’ (AQT) Intelligent Quantum Networks and Technologies  (IN-Q-NET) research program that supports the FQNET, CQNET, IEQNET and other research quantum netwroking and communications testbeds. 
R.V., N.L., L.N., N.S., M.S., G.F.,  S.X. and A.M.  acknowledge partial support from the U.S. Department of Energy, Office of Science, High Energy Physics, QuantISED program grant, under award number DE-SC0019219. A.M. is supported  in part by the JPL President and Director’s Research and Development Fund (PDRDF). C.P. further acknowledges partial support from Fermilab's LDRD. 
J.A. acknowledges support by a NASA Space Technology Research Fellowship. 
Part of the research was carried out at the Jet Propulsion Laboratory, California Institute of Technology, under a contract with the National Aeronautics and Space Administration (80NM0018D0004).
We thank   Jason Trevor (Caltech Lauritsen Laboratory for High Energy Physics and INQNET Laboratory for QST),  Vikas Anant (PhotonSpot), Aaron Miller (Quantum Opus), Inder Monga and his ESNET and QUANT-NET groups at LBNL, the groups of Daniel Oblak and Christoph Simon at the University of Calgary, the group of Marko Loncar  at Harvard, Artur Apresyan and the HL-LHC USCMS-MTD Fermilab group; Marco Colangelo (MIT);  We acknowledge the enthusiastic support of the Kavli Foundation on funding QS\&T workshops and events and the Brinson Foundation support for students working at FQNET and CQNET.

\bibliographystyle{IEEEtran}
\raggedright
\bibliography{Bibliography}

\begin{thebibliography}{10}
\providecommand{\url}[1]{#1}
\csname url@rmstyle\endcsname
\providecommand{\newblock}{\relax}
\providecommand{\bibinfo}[2]{#2}
\providecommand\BIBentrySTDinterwordspacing{\spaceskip=0pt\relax}
\providecommand\BIBentryALTinterwordstretchfactor{4}
\providecommand\BIBentryALTinterwordspacing{\spaceskip=\fontdimen2\font plus
\BIBentryALTinterwordstretchfactor\fontdimen3\font minus
  \fontdimen4\font\relax}
\providecommand\BIBforeignlanguage[2]{{%
\expandafter\ifx\csname l@#1\endcsname\relax
\typeout{** WARNING: IEEEtran.bst: No hyphenation pattern has been}%
\typeout{** loaded for the language `#1'. Using the pattern for}%
\typeout{** the default language instead.}%
\else
\language=\csname l@#1\endcsname
\fi
#2}}

\bibitem{valivarthi2020teleportation}
R.~Valivarthi, S.~I. Davis, C.~Pe{\~n}a, S.~Xie, N.~Lauk, L.~Narv{\'a}ez, J.~P.
  Allmaras, A.~D. Beyer, Y.~Gim, M.~Hussein, \emph{et~al.}, ``Teleportation
  systems toward a quantum internet,'' \emph{PRX Quantum}, vol.~1, no.~2, p.
  020317, 2020.

\bibitem{valivarthi2016quantum}
R.~Valivarthi, Q.~Zhou, G.~H. Aguilar, V.~B. Verma, F.~Marsili, M.~D. Shaw,
  S.~W. Nam, D.~Oblak, W.~Tittel, \emph{et~al.}, ``Quantum teleportation across
  a metropolitan fibre network,'' \emph{Nature Photonics}, vol.~10, no.~10, pp.
  676--680, 2016.

\bibitem{alshowkan_2022}
M.~Alshowkan, P.~G. Evans, B.~P. Williams, N.~S. Rao, C.~E. Marvinney, Y.-Y.
  Pai, B.~J. Lawrie, N.~A. Peters, and J.~M. Lukens, ``Advanced architectures
  for high-performance quantum networking,'' \emph{Journal of Optical
  Communications and Networking}, vol.~14, no.~6, p. 493–499, 2022.

\bibitem{chung_2021}
J.~Chung, G.~Kanter, N.~Lauk, R.~Valivarthi, W.~Wu, R.~R. Ceballos, C.~Peña,
  N.~Sinclair, J.~Thomas, S.~Xie, and et~al., ``Illinois express quantum
  network (ieqnet): Metropolitan-scale experimental quantum networking over
  deployed optical fiber,'' \emph{Quantum Information Science, Sensing, and
  Computation XIII}, vol. 11726, Apr 2021.

\bibitem{tanaka08}
\BIBentryALTinterwordspacing
A.~Tanaka, M.~Fujiwara, S.~W. Nam, Y.~Nambu, S.~Takahashi, W.~Maeda, K.~ichiro
  Yoshino, S.~Miki, B.~Baek, Z.~Wang, A.~Tajima, M.~Sasaki, and A.~Tomita,
  ``Ultra fast quantum key distribution over a 97 km installed telecom fiber
  with wavelength division multiplexing clock synchronization,'' \emph{Opt.
  Express}, vol.~16, no.~15, pp. 11\,354--11\,360, Jul 2008. [Online].
  Available: \url{http://www.opticsexpress.org/abstract.cfm?URI=oe-16-15-11354}
\BIBentrySTDinterwordspacing

\bibitem{lessing_2017}
M.~Lessing, H.~S. Margolis, C.~T. Brown, and G.~Marra, ``Frequency comb-based
  time transfer over a 159 km long installed fiber network,'' \emph{Applied
  Physics Letters}, vol. 110, no.~22, p.~=, May 2017.

\bibitem{chen_2017}
X.~Chen, Y.~Cui, X.~Lu, C.~Ci, X.~Zhang, B.~Liu, H.~Wu, T.~Tang, K.~Shi,
  Z.~Zhang, and et~al., ``High-precision multi-node clock network
  distribution,'' \emph{Review of Scientific Instruments}, vol.~88, no.~10, Oct
  2017.

\bibitem{wu_2019}
R.~Wu, J.~Lin, T.~Jiang, C.~Liu, and S.~Yu, ``Stable radio frequency transfer
  over fiber based on microwave photonic phase shifter,'' \emph{Optics
  Express}, vol.~27, no.~26, p. 38109–38115, Dec 2019.

\bibitem{thomas_2021}
J.~M. Thomas, G.~S. Kanter, K.~F. Lee, and P.~Kumar, ``Coexistence of entangled
  and classical light over 45 km of installed fiber,'' \emph{Frontiers in
  Optics + Laser Science 2021}, 2021.

\bibitem{Valivarthi:2022vni}
R.~Valivarthi \emph{et~al.}, ``{Picosecond synchronization system for quantum
  networks},'' 3 2022.

\bibitem{frohlich_2015}
B.~Fröhlich, J.~F. Dynes, M.~Lucamarini, A.~W. Sharpe, S.~W.-B. Tam, Z.~Yuan,
  and A.~J. Shields, ``Quantum secured gigabit optical access networks,''
  \emph{Scientific Reports}, vol.~5, no.~1, 2015.

\bibitem{QIB:DOE2020-21}
``{DOE's Quantum Internet Blueprint},''
  \url{https://www.energy.gov/sites/prod/files/2020/07/f76/QuantumWkshpRpt20FINAL_Nav_0.pdf}.

\bibitem{polatis}
``Polatis,''
  \url{https://www.polatis.com/series_6000_multimode_switch_all-optical_switching.asp}.

\bibitem{ieqnet_design}
J.~Chung \emph{et~al.}, ``{IEQNET: Design and Implementation of a Quantum
  Metropolitan Area Network},'' \emph{Submitted to IEEE-TQE}, 2022.

\bibitem{tektronix}
``Tektronix,'' \url{https://https://www.tek.com/arbitrary-waveform-generator/}.

\bibitem{ixblue}
``Ixblue,'' \url{https://www.ixblue.com/north-america/store/dr-ve-10-mo/}.

\bibitem{loudon2000}
R.~Loudon, \emph{The quantum theory of light}.\hskip 1em plus 0.5em minus
  0.4em\relax OUP Oxford, 2000.

\bibitem{korzh2020lowjitter}
\BIBentryALTinterwordspacing
B.~Korzh, Q.-Y. Zhao, J.~P. Allmaras, S.~Frasca, T.~M. Autry, E.~A. Bersin,
  A.~D. Beyer, R.~M. Briggs, B.~Bumble, M.~Colangelo, G.~M. Crouch, A.~E. Dane,
  T.~Gerrits, A.~E. Lita, F.~Marsili, G.~Moody, C.~Pe{\~{n}}a, E.~Ramirez,
  J.~D. Rezac, N.~Sinclair, M.~J. Stevens, A.~E. Velasco, V.~B. Verma, E.~E.
  Wollman, S.~Xie, D.~Zhu, P.~D. Hale, M.~Spiropulu, K.~L. Silverman, R.~P.
  Mirin, S.~W. Nam, A.~G. Kozorezov, M.~D. Shaw, and K.~K. Berggren,
  ``Demonstration of sub-3 ps temporal resolution with a superconducting
  nanowire single-photon detector,'' \emph{Nature Photonics}, vol.~14, no.~4,
  pp. 250--255, Apr 2020. [Online]. Available:
  \url{https://doi.org/10.1038/s41566-020-0589-x}
\BIBentrySTDinterwordspacing

\bibitem{ASCR:IEQNET2019}
``{IEQNET},''
  \url{https://science.osti.gov/-/media/ascr/pdf/funding/2019/ASCR_Quantum_Computing_and_Quantum_Networking_Awards_FY-2019.pdf}.

\end{thebibliography}


\end{document}